\documentclass[conference,10pt,letter]{IEEEtran}

\usepackage{amsmath, amsfonts}
\usepackage{cite}
\usepackage{algorithm}
\usepackage{algpseudocode}
\usepackage{enumerate}
\usepackage{color}
\usepackage{setspace}
\usepackage{url}
\usepackage{balance}
\usepackage{multirow}  
\usepackage{pbox}  
\usepackage{flushend}

\usepackage[pdftex]{graphicx}
\graphicspath{{./}}%
\DeclareGraphicsExtensions{.jpg,.png,.pdf}
\usepackage[font={footnotesize}]{caption}
\usepackage[caption=false,font=footnotesize]{subfig}

\newcommand{\bi}{\begin{itemize}}	
\newcommand{\ei}{\end{itemize}}
\newcommand{\bn}{\begin{enumerate}}	
\newcommand{\en}{\end{enumerate}}
\newcommand{\bc}{\begin{center}}
\newcommand{\ec}{\end{center}}
\newcommand{\be}{\begin{equation}}
\newcommand{\ee}{\end{equation}}
\newcommand{\ben}{\begin{equation*}}
\newcommand{\een}{\end{equation*}}
\newcommand{\beqa}{\begin{eqnarray}}
\newcommand{\eeqa}{\end{eqnarray}}
\newcommand{\btabu}{\begin{tabular}}
\newcommand{\etabu}{\end{tabular}}

\newcommand{\mb}{\mathbf}

\begin{document}
\title{Fingerprinting-Based Positioning in Distributed Massive MIMO Systems}

\author{\IEEEauthorblockN{Vladimir Savic and Erik G. Larsson \vspace{0.5em}}
\IEEEauthorblockA{$^1$Dept. of Electrical Engineering (ISY), Link\"{o}ping University, Link\"{o}ping, Sweden\\
Emails: vladimir.savic@liu.se, erik.larsson@isy.liu.se}
}
\maketitle

\begin{abstract}
Location awareness in wireless networks may enable many applications such as emergency services, autonomous driving and geographic routing. Although there are many available positioning techniques, none of them is adapted to work with massive multiple-in-multiple-out (MIMO) systems, which represent a leading 5G technology candidate. In this paper, we discuss possible solutions for positioning of mobile stations using a vector of signals at the base station, equipped with many antennas distributed over deployment area. Our main proposal is to use fingerprinting techniques based on a vector of received signal strengths. This kind of methods are able to work in highly-cluttered multipath environments, and require just one base station, in contrast to standard range-based and angle-based techniques. We also provide a solution for fingerprinting-based positioning based on Gaussian process regression, and discuss main applications and challenges.
\end{abstract}

\section{Introduction}\label{sec:intro}\footnote{Copyright (c) 2015 IEEE. Personal use of this material is permitted. However, permission to use this material for any other purposes must be obtained from the IEEE by sending a request to pubs-permissions@ieee.org. The original version of this manuscript is published in IEEE Proc. of 82nd Vehicular Technology Conference (VTC2015-Fall).}
Massive (or large-scale) MIMO system, a leading 5G technology candidate, was conceived in \cite{Mar2010} and relies on using a large number of base station (BS) antennas to serve large numbers of mobile stations (MS) simultaneously. The main advantages of massive MIMO are that: i) it can provide uniformly good service to everyone in the cell, and ii) it can boost spectral efficiencies compared to current standards by a factor of ten or more.  Due to channel hardening, scheduling and power control can be performed on a slow time-scale and subcarrier-independently. Massive MIMO relies on time-division-duplex operation and reciprocity. Calibration for transmit-receive reciprocity imbalances in the hardware is necessary, but the array elements need not be mutually phase-calibrated. Tutorial papers that describe the operation of massive MIMO and its performance in more detail include \cite{Mar2010,LTEM:14:CM}. A potential improvement of massive MIMO could be achieved by distributing BS antennas over larger space, e.g., over roof of the building. This type of massive MIMO, also known as \textit{distributed} (or \textit{cell-free}) massive MIMO (DM-MIMO) \cite{Joung2014,Liu2014,Truong2013,Ngo2015}, is especially beneficial for positioning due to the better spatial diversity. Therefore, in this paper, we focus on positioning using DM-MIMO.

In DM-MIMO, the base stations receive on the uplink large amounts of data and potentially, this data can be exploited for positioning services. Compared to conventional systems (even MIMO), potentially more accurate positioning is possible owing to the increased size of the measurement vector. At the same time, the amount of data to be processed is huge, so the design of fast positioning algorithms is an important challenge. In this paper, we discuss some of the opportunities for accurate and efficient positioning with DM-MIMO, and outline some basic positioning techniques that could be used to process the data in order to obtain position estimates. More specifically, we discuss range-based, angle-based, and fingerprinting (FP) techniques. Our main proposal is to use FP techniques based on a vector of received signal strengths (RSS), since they are able to work in highly-cluttered multipath environments, and require just one BS. 

The remainder of this paper is organized as follows. In Section~\ref{sec:posMIMO}, we discuss positioning techniques that can be used with DM-MIMO. In Section \ref{sec:FPpos}, FP-based positioning using Gaussian process regression (GPR) is described. In Section \ref{sec:numex}, a numerical example is provided. A discussion on potential applications and the main challenges is provided in Section \ref{sec:opport}. Finally, conclusions are provided in Section~\ref{sec:conc}.

\section{Possible positioning techniques}\label{sec:posMIMO}
\begin{figure}[!th]
\centerline{
\subfloat[]{\includegraphics[width=0.97\columnwidth]{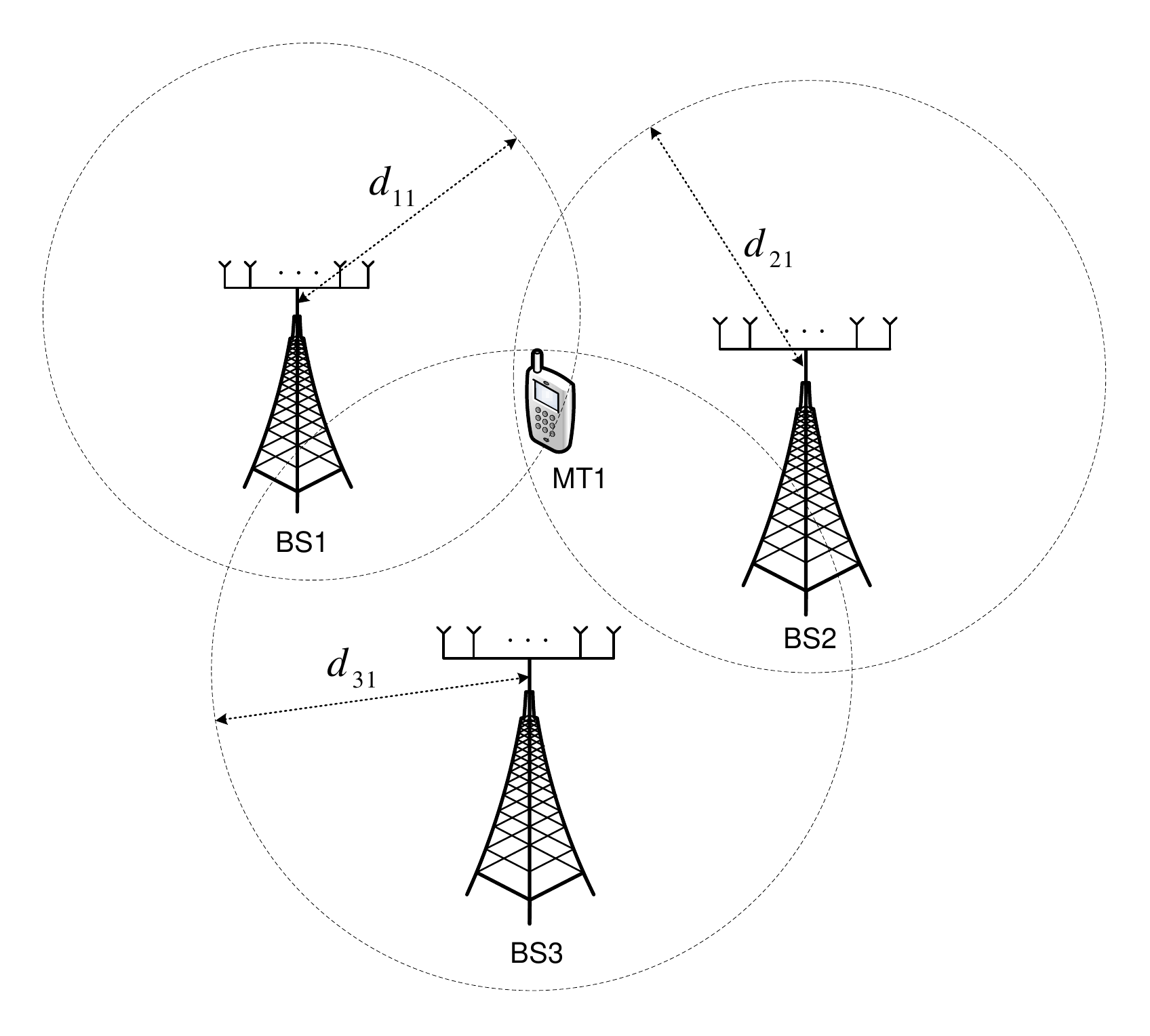}}\label{fig:mmimo-tril}
}
\centerline{
\subfloat[]{\includegraphics[width=0.97\columnwidth]{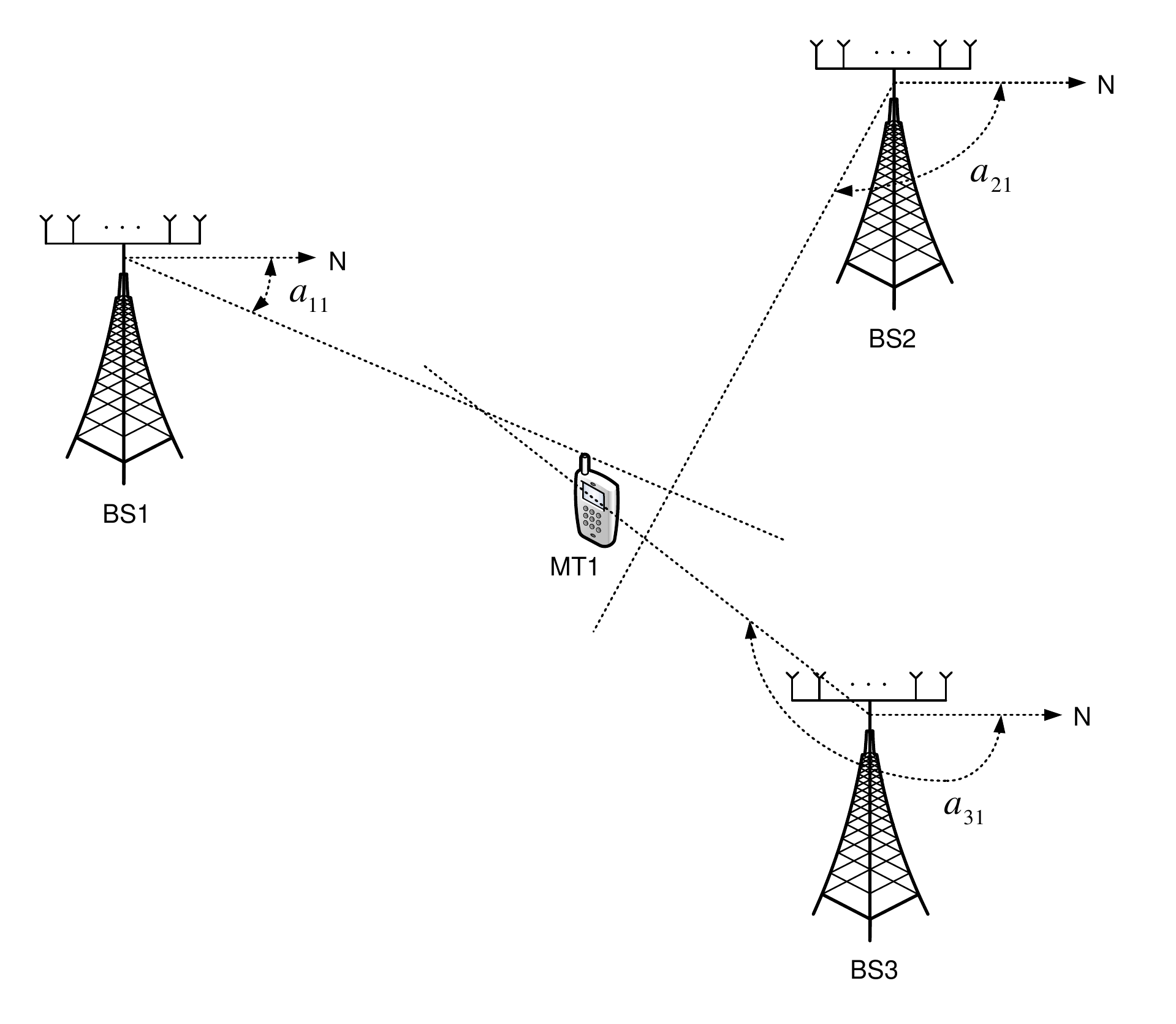}}\label{fig:mmimo-trian}
}
\centerline{
\subfloat[]{\includegraphics[width=0.92\columnwidth]{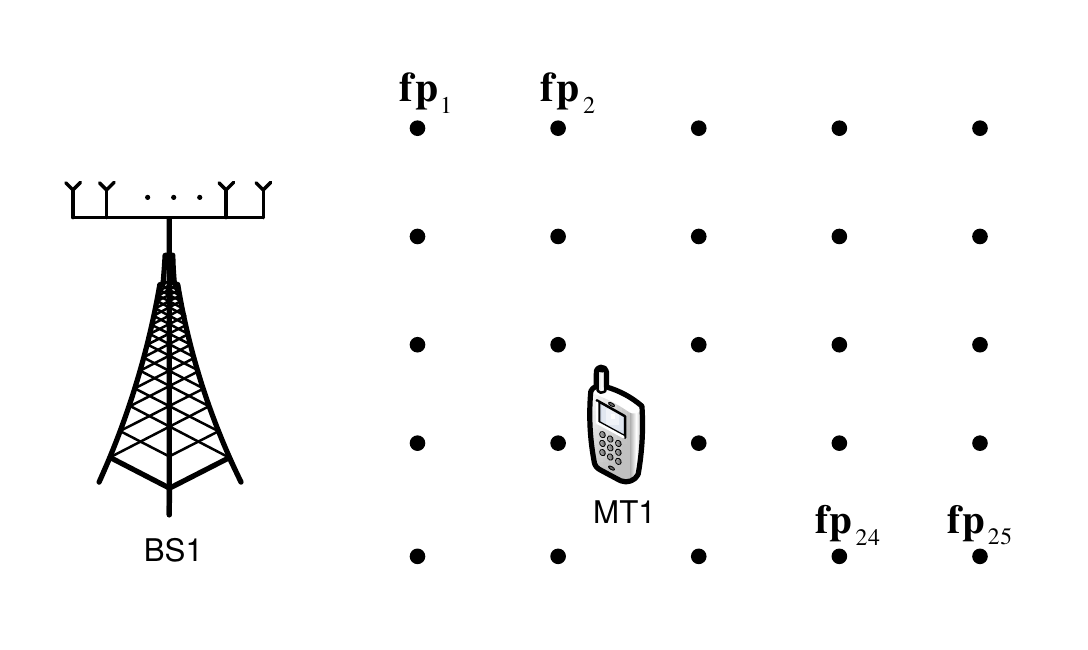}}\label{fig:mmimo-fp}
}
\caption{Illustration of positioning with DM-MIMO: (a) trilateration, (b) triangulation, (c) FP. The estimated ranges are denoted by $d_{11}$, $d_{21}$, $d_{31}$, the estimated angles by $a_{11}$, $a_{21}$, $a_{31}$, and the fingerprints by $\mb{fp}_1$,..., $\mb{fp}_{25}$.}
\label{fig:mmimoPos}
\end{figure}

Although all available positioning techniques \cite{Gustafsson2005, Mao2007a} for single-antenna systems may be used with DM-MIMO, most of them are not preferable. Generally, the positioning techniques can be divided into the four main classes: i) proximity-based, ii) angle-based, iii) range-based, and iv) fingerprinting-based. The \textit{proximity-based} (or \textit{cell-id}) techniques can be immediately ruled out since many BSs would be required, which is not available, for example, in a suburban environment. The \textit{angle-based} techniques may be feasible since the angle of arrival can be measured using the antenna array on the BS (i.e., using phase interferometry \cite{Mao2007a}), which can be then used to find the position by triangulation. However, this approach may fail in case of non-line-of-sight (NLOS), and lead to large positioning errors. The third option is to obtain a range between the MT and at least three BSs (or one BS, if its antennas are separated enough), and then find the MT position by trilateration. This can be done either by using \textit{time-of-arrival} (TOA), or by using \textit{received signal strength} (RSS) measurements. Since TOA estimation requires a high signal bandwidth, it is not suitable for cellular networks (e.g., the bandwidth of 20 MHz would lead to a ranging resolution of 15 m). The RSS measurements may be useful in outdoor non-urban environments (in which the path-loss is expected to monotonically decrease with the distance), but in most other environments would lead to very coarse range estimates. These problems can be solved using \textit{fingerprinting} (FP) techniques, as will be argued below. The illustration of positioning techniques for DM-MIMO is shown in Fig. \ref{fig:mmimoPos}. 

In FP-based positioning, the goal is to \textit{directly} find the MT's position using an online measurement and a set of training samples, also known as \textit{fingerprints}. The fingerprints consist of the measurements obtained at known positions within the deployment area, and they can be obtained either offline or online. Assuming that the input measurement is high-dimensional (as in DM-MIMO), one BS with fixed (even unknown) position is enough for MT positioning. Moreover, the position estimates are much less sensitive to multipath, since multipath can be captured during the training phase. On the other hand, the main drawbacks of FP methods are: i) higher computational complexity, and ii) possible performance degradation in dynamic environments. While the former problem is not expected to be a big issue with nowadays computational recourses, the latter problem can be resolved (or at least, reduced) with online training. Therefore, although not ideal, we believe the FP-based positioning could be the right choice for positioning with DM-MIMO. In what follows, we provide a brief review of few well-known FP methods:
\bi
\item \textit{$\kappa$ Nearest Neighbors ($\kappa$NN)} \cite{Ni2003}: This method assumes that there are many reference points at which the fingerprints (vectors of RSS) are obtained, and one target (MT) to be located. The position of the object is found as a weighted average of the closest $\kappa$ reference positions. The distance between the target and the reference points is measured in the signal space (e.g., Euclidean distance between corresponding RSS vectors), and then the closest $\kappa$ positions are chosen ($\kappa$ is chosen empirically so as to minimize positioning error). The weights are usually equal to the squared inverse of this distance, but other options are also possible. $\kappa$NN is able to provide very good performance if the reference points are uniformly distributed across the deployment area.

\item \textit{Support Vector Machines (SVM)} \cite{Brunato2005}: SVM is a supervised machine learning technique based on convex optimization. It can be used for regression problem, such as FP-based positioning, in which the goal is to directly estimate the MT position from an online measurement and a set of training points (fingerprints). Since it is a nonparametric approach, it is able to model any arbitrary nonlinear relationship (assuming that there are enough fingerprints). Moreover, this method is much cheaper than other nonparametric methods since the estimation is based on a subset of the training samples (known as \textit{support vectors}). 

\item \textit{Gaussian Process Regression (GPR}) \cite{Perez-Cruz2013}: GPR is a Bayesian nonparametric approach for regression. With this method, an unknown nonlinear function is assumed to be random, and to follow a Gaussian process (GP). In contrast to SVM and $\kappa$NN, GPR is able to provide probabilistic output, i.e., the posterior distribution of the MT position, given an online measurement and a set of fingerprints. On the other hand, it is much more computationally complex since, at least in its original form, it uses all fingerprints for estimation. More details about this method is available in the following section.
\ei

\section{Fingerprinting-based positioning}\label{sec:FPpos}
The goal is to locate MTs using the received signal at BS. Therefore, we consider the uplink of a multi-user DM-MIMO system with $M$ antennas on the BS, and $K$ single-antenna MTs. Assuming that MTs simultaneously transmit $K$ symbols, $\mb{s}=(s_1,\ldots,s_K)^T$, the BS receives:
\be\label{eq:rsignal}
\mb{y}=\sqrt{\rho}\mb{G}\mb{s}+\mb{w}
\ee
where $\mb{G}$ is the $M \times K$ channel matrix, with entries $g_{k,m}=h_{k,m}\sqrt{\beta_{k,m}}$ ($k=1,\ldots,K$, $m=1,\ldots,M$); $\mb{w}=(w_1,\ldots,w_M)$ is the noise vector (typically, $w_m \sim \mathcal{CN}(0,1)$) and $\rho$ is the SNR seen at MT, assuming that the input signal power is normalized (i.e., $E(\lVert \mb{s}\rVert^2)=1$); $h_{k,m}$ and $\beta_{k,m}$ are, respectively, small-scale and large-scale fading between $k$th MT and $m$th BS antenna. We assume here that the large-scale fading (shadowing) is constant with respect to frequency, and the small-scale fading follows a complex Gaussian distribution, i.e., $h_{k,m}\sim \mathcal{CN}(0,1)$ (i.e., $|h_{k,m}|$ follows Rayleigh distribution). Note that, assuming OFDM modulation, $\mb{y}=(y_1,\ldots,y_M)$ represents the received signal of one subcarrier. The number of signals that BS receives within one coherence interval is $N_c=N_{slot}N_{sub}$, where $N_{slot}$ is the number of transmitted symbols, and $N_{sub}$ is the number of subcarriers. To keep the notation uncluttered, the indexes of subcarriers and symbols are omitted.

As we argued in the previous section, the most appropriate information for positioning is the vector of RSS, given by $\lVert \mb{y}\rVert^2$ for one subcarrier. However, since this signal includes the combined power from $K$ terminals, it is not appropriate for positioning. Therefore, we need the terminals to transmit an orthogonal set of pilot signals \cite{Ngo2015}, that are already used for signal detection. Moreover, to ensure that one BS is sufficient for positioning, the BS antennas should be well separated. That means that DM-MIMO is required for this kind of problem.\footnote{With collocated massive MIMO, multiple BSs (at least, 3) would be required.} Another problem is variation of RSS due to the small-scale fading, which can be reduced by averaging the received power over all subcarriers (if not enough, the averaging over multiple symbols would be also required). This procedure, also known as \textit{channel hardening}, will ensure that, for Rayleigh channel, RSS solely depends on the path-loss between the BS and the MS. Taking all assumptions into account, the RSS between the $k$th MT and $m$th BS antenna is proportional to large-scale fading, $\beta_{k,m}$. One way to model this power is by using log-distance path-loss model:
\be\label{eq:powerBSMS}
p_{k,m} \propto \log{\beta_{k,m}}=p^0_{k,m}-10n_p\log{(d_{k,m}/d_{k,m}^0)}+v
\ee
where $d_{k,m}$ is the distance between $k$th MT and $m$th BS antenna, $p^0_{k,m}$ is the reference power measured at distance $d_{k,m}^0$, $n_p$ is the path-loss exponent (typically, $0 < n_p < 6$ depending on environment and the range), and $v \sim \mathcal{N}(0,\sigma^2_p)$ is the shadowing noise. For urban environments, a multiple-slope model for path-loss is typically used. Note that this model is simplified, i.e., the power is not necessarily decreasing with the distance, and the shadowing is not necessarily Gaussian. However, the following algorithm will be also valid for any other model, as long as RSS is variable over space (i.e., each position has an unique fingerprint).

Given the RSS vector $\mb{p}=(p_{k,1},\ldots,p_{k,M})$, our goal is to find the position of the MT in the 2-dimensional plane, denoted by $(x_{1},x_{2})$. Since we focus on a single MT, the index $k$ is omitted. Therefore, we have to solve the following regression problem:
\be\label{eq:fpRegress} 
x_{i}=f_{i}(\mb{p})+\nu_{i},~~(i=1,2)
\ee 
where $f_{i}(\cdot)$ is a nonlinear function of the input vector, and $\nu_{i}$ is a Gaussian random variable ($\nu_{i} \sim \mathcal{N}(0,\sigma^2_{\nu,i})$) that represents the error. The problem can be solved using some of the techniques described in the previous section, but we focus on GPR since it provides a probabilistic output. In that case, the function $f_{i}(\cdot)$ is assumed to be \textit{random}, and follows a GP: $f_{i}(\cdot) \sim \mathcal{GP}({\bf{m}}_{i},{\bf{C}}_{i})$ where ${\bf{m}}_{i}$ is a mean function (typically, ${\bf{m}}_{i}=0$) and ${\bf{C}}_{i}$ is a covariance function (also known as a \textit{kernel} matrix). The kernel matrix is used to model the correlation between output samples as a function of the input samples. Although there are many possible options \cite{Rasmussen2006}, a widely used kernel is a weighted sum of the squared-exponential and the linear function:
\be\label{eq:kernelSqExp}
c_i(\mb{p}_{l},\mb{p}_{j}) = {\theta _0}{e^{ - {\theta _1}{{\left\| {\mb{p}_{l} - \mb{p}_{j}} \right\|}^2}}} + {\theta _2}{\bf{p}}_l^T{{\bf{p}}_j}
\ee
where $c_i(\mb{p}_{l},\mb{p}_{j})$ is an entry of the $\mb{C}_i$, and $\mb{p}_{l},\mb{p}_{j}$ are any two measurement samples ($l,j=1,\ldots,L$). The hyperparameters $\boldsymbol\theta=(\theta _0,\theta _1,\theta _2)$, along with $\sigma_{\nu,i}$, can be estimated from the training data. The intuition behind this kernel is that the correlation between the output samples should be higher if the Euclidean distance between the corresponding input samples is smaller. 

Assuming that we have available a set of \textit{i.i.d.} training samples (fingerprints) ${{\cal T}_{L,i}} = \{x_{i,l},\mb{p}_{l}\} _{l = 1}^L$ and a single test measurement $\mb{p}$, we would like to determine the
posterior density of the position, i.e., $p(x_{i}|\mb{p},\mathcal{T}_{L,i})$ ($i=1,2$). It can be shown \cite{Perez-Cruz2013} that this distribution is Gaussian, with the following mean and the variance:
\be\label{eq:postMean}
{\mu _{GPR,i}} = {{\bf{c}}_i^T}{({\bf{C}}_i + \sigma_{\nu,i} ^2{{\bf{I}}_L})^{ - 1}}{{\bf{x}}_{i}}
\ee
\be\label{eq:postVar}
\sigma _{GPR,i}^2 = \sigma _{\nu,i} ^2 + c_i({\bf{p}},{\bf{p}}) - {{\bf{c}}_i^T}{({\bf{C}}_i + \sigma _{\nu,i} ^2{{\bf{I}}_L})^{ - 1}}{\bf{c}}_i
\ee
where we define the vectors: ${\bf{c}}_i = {\left( {c_i({{\bf{p}}_1},{\bf{p}}),\, \ldots ,c_i({{\bf{p}}_L},{\bf{p}})} \right)^T}$ and ${{\bf{x}}_{i}} = ({x_{1,i}}, \ldots ,\,{x_{L,i}})^T$. Therefore, the MMSE estimate, of the $i$th coordinate of the position, is given by $\hat{x}_{i}={\mu _{GPR,i}}$, and the remaining uncertainty by the variance $\sigma _{GPR,i}^2$. Thus, GPR is capable of providing complete statistical information of the MT position. Another important characteristic of GPR is that $\sigma _{GPR,i}$ is small in the areas where the training samples lie, and large in the areas with no (or few) training samples. 

Regarding computational complexity of GPR, the offline phase (based on training samples) is dominated by computation of the inverse of the matrix ${{\bf{C}}_i + \sigma_{\nu,i} ^2{{\bf{I}}_L}}$, which requires $\mathcal{O}(L^{\eta})$ operations, where $2 \le \eta \le3$ depends on the type of approximation. Assuming that this matrix is stored into memory, the online phase of GPR can be achieved in $\mathcal{O}(L^2)$ operations.

\begin{figure}[!t]
\centerline{
\subfloat[]{\includegraphics[width=1\columnwidth]{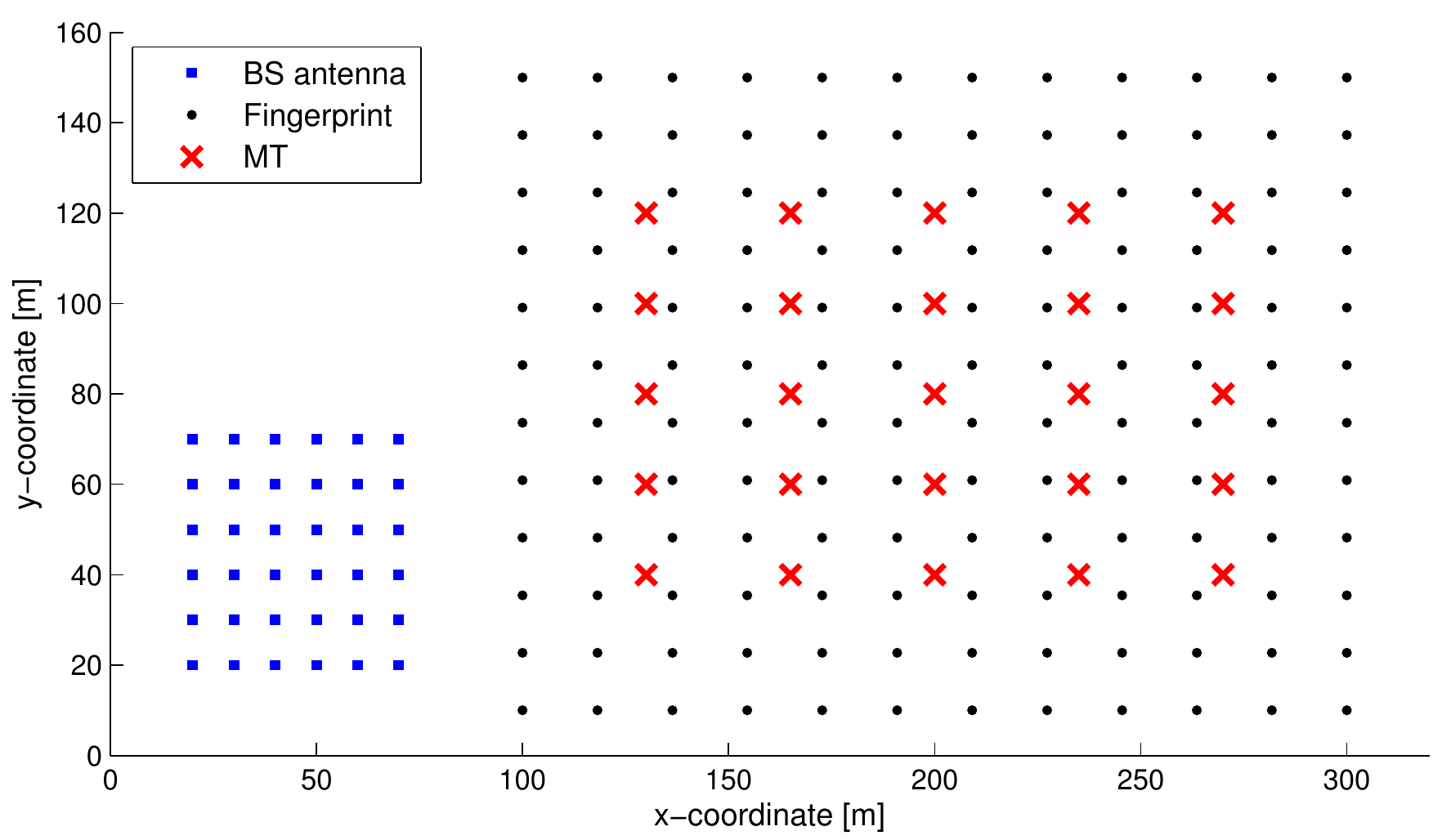}}\label{fig:DMMIMOscen1}
}
\centerline{
\subfloat[]{\includegraphics[width=1\columnwidth]{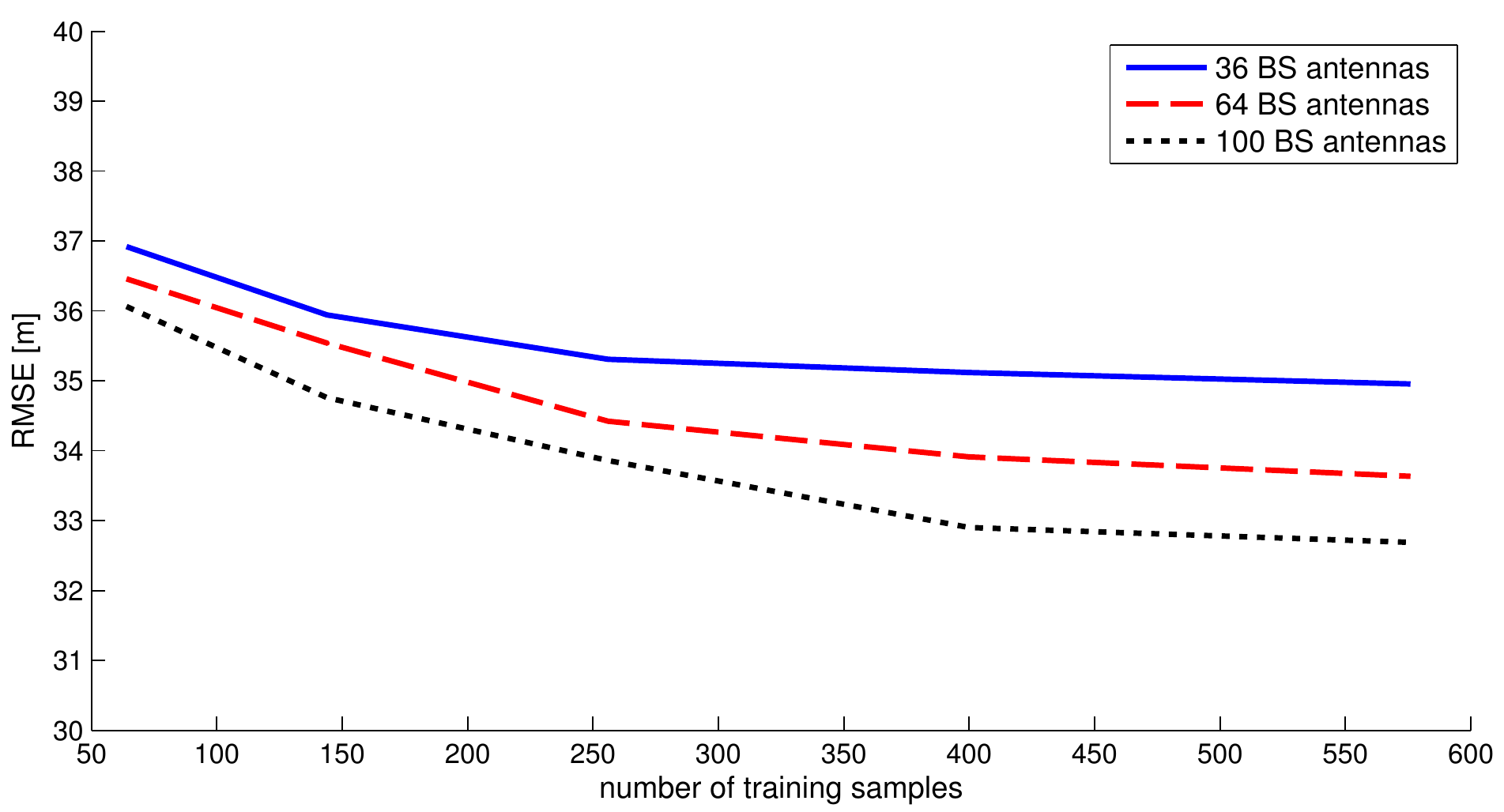}}\label{fig:RMSEscen1}
}
\centerline{
\subfloat[]{\includegraphics[width=1\columnwidth]{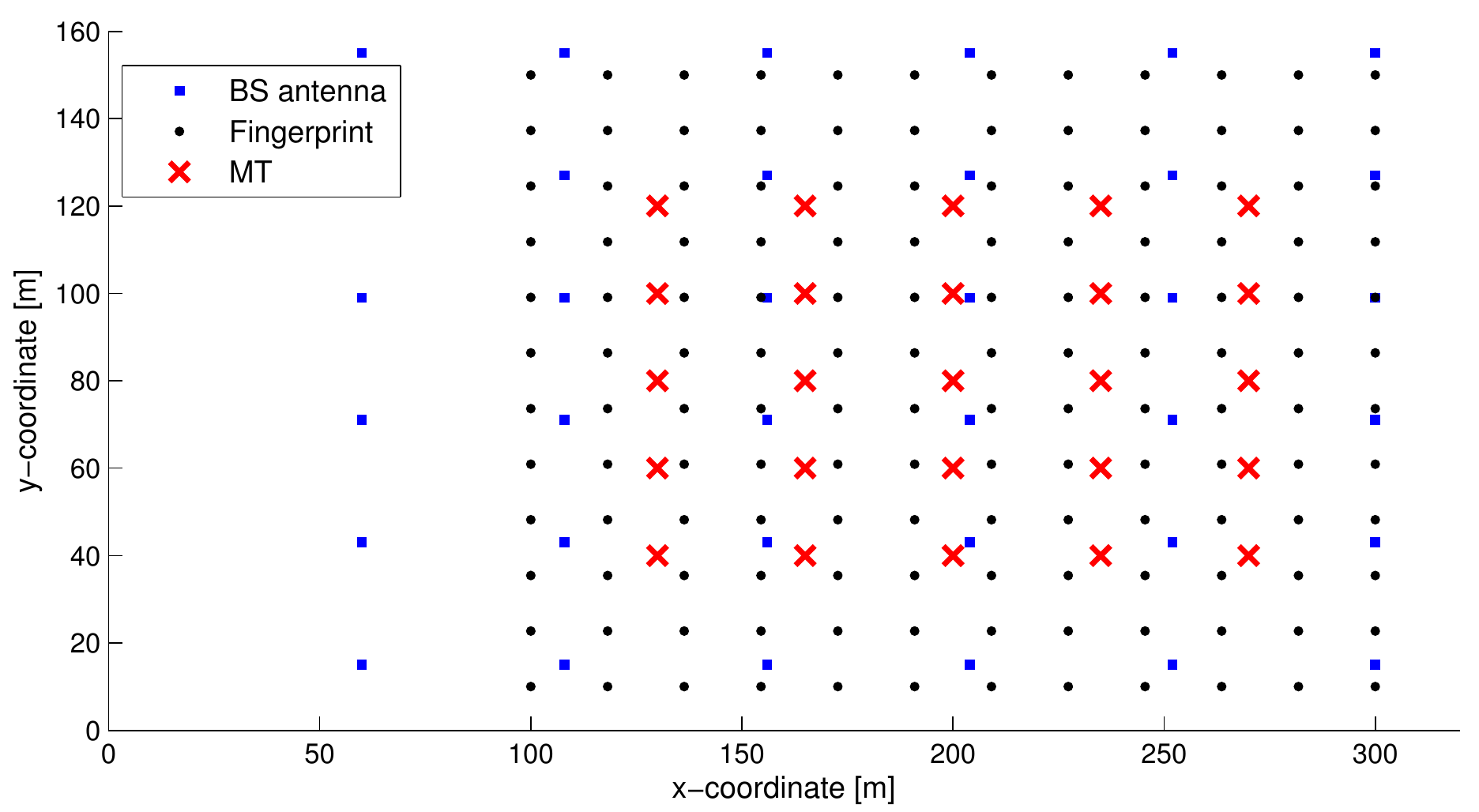}}\label{fig:DMMIMOscen2}
}
\centerline{
\subfloat[]{\includegraphics[width=1\columnwidth]{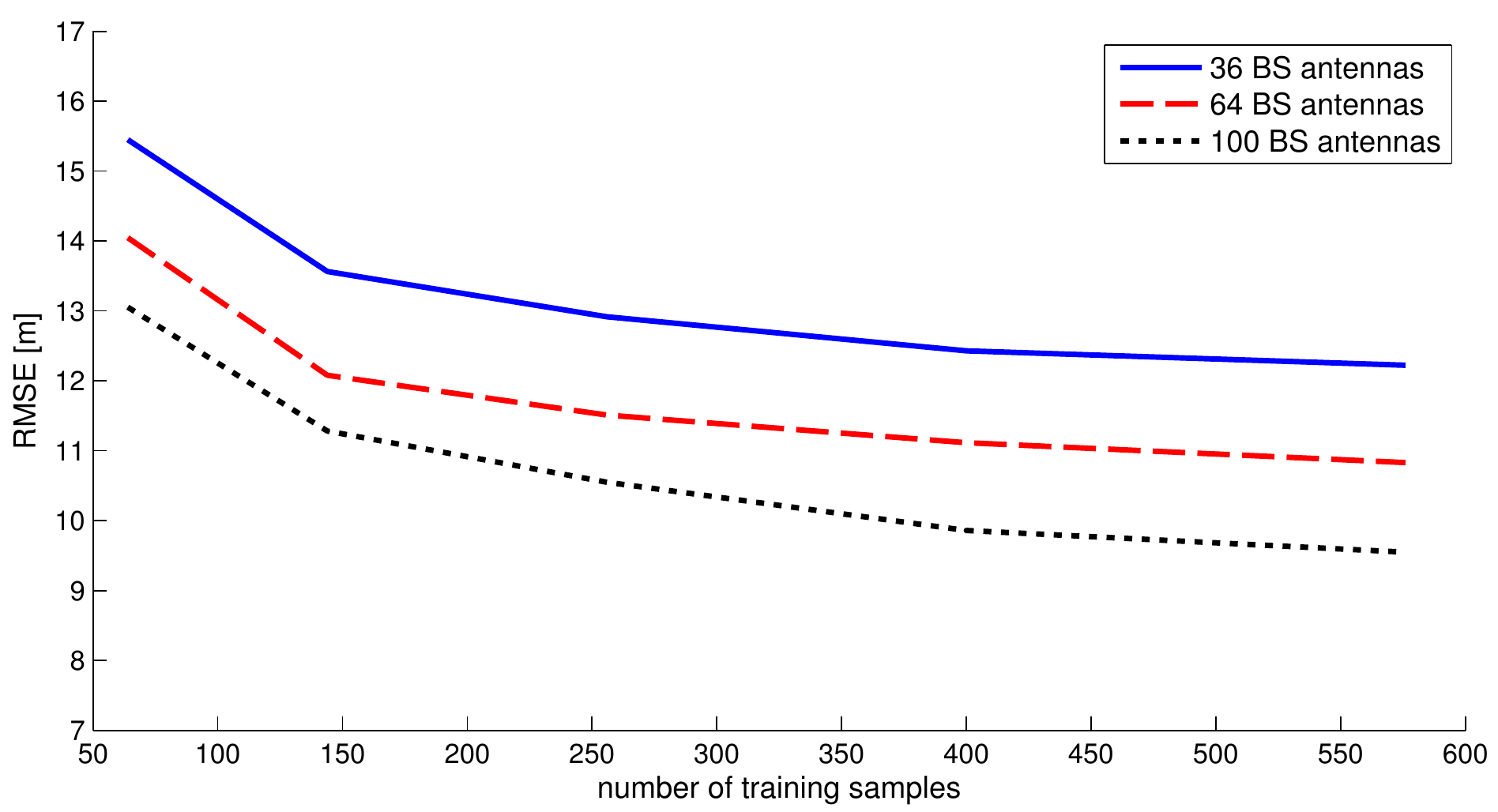}}\label{fig:RMSEscen2}
}
\caption{(a), (c) The deployment of BS, MTs, and the fingerprints; (b), (d) RMSE of the MT positions for the deployment from (a) and (c), respectively.}
\label{fig:MIMOresultsGPR}
\end{figure}

\section{Numerical example}\label{sec:numex}
We consider the scenario with one BS with $M=36,64,100$ antennas, and $K=25$ MTs. The path-loss exponent is set to $n_p=0$ for $0<d_{k,m}<10~\mathrm{m}$, $n_p=2$ for $10~\mathrm{m}<d_{k,m}<50~\mathrm{m}$, and $n_p=4$ for $d_{k,m}>50~\mathrm{m}$ ($k=1,\ldots,K$, $m=1,\ldots,M$). The shadowing variance is $\sigma_p=5$ dB, and the reference power is $p^0_{k,m}(d_{k,m}^0=10~\mathrm{m})=0$ dB. Note that, thanks to the channel hardening, the parameters for the small-scale model in \eqref{eq:rsignal} are not required for this numerical analysis. The BS antennas, MTs and fingerprints are distributed in grid configurations, as shown in Fig. \ref{fig:MIMOresultsGPR}a and Fig. \ref{fig:MIMOresultsGPR}c. We will analyse the root-mean-square-error (RMSE) of the MT positions, as function of the number of training samples $L$, and the number of the BS antennas $M$. The results are averaged over 200 Monte Carlo runs, and in each of them the average is performed over $K$ MTs. 

As can be seen in Fig. \ref{fig:MIMOresultsGPR}b and Fig. \ref{fig:MIMOresultsGPR}d, RMSE error is decreasing with the number of training samples. However, the gain is small for $L>400$ since the fingerprints become more correlated as their density is increasing. We also note that increasing the number of BS antennas leads to an improvement in performance, which motivates the use of DM-MIMO. By comparing Fig. \ref{fig:MIMOresultsGPR}b and Fig. \ref{fig:MIMOresultsGPR}d, in which two different BS configurations have been used, we can see that there is a substantial difference in the performance. In general, spreading the antennas over larger area increases position accuracy, but it also increases communication overhead due to the backhauling of the data over larger distances.

\section{Opportunities and challenges}\label{sec:opport}

Location information in massive MIMO systems may enable many applications for future (5G) networks \cite{Taranto2014}. We discuss below the most relevant ones.
\bi
\item \textit{GPS-free position awareness}: Although the Global Positioning System (GPS) provides relatively accurate position estimates (5 to 15 meters), it is not suitable for indoor, dense urban and other NLOS environments. Moreover, equipping every user with a GPS is more expensive than a local positioning system (LPS) based on radio signals. DM-MIMO would provide, not only the location awareness, but also increased robustness against single-antenna failures. This is crucial for location-based services such as emergency calls, and weather forecasts.

\item \textit{Increasing spectrum efficiency}: Position awareness could significantly increase spectral efficiency since the frequency slots may be re-used for the MTs that are well separated. This idea is already used in cognitive networks, where users may choose the frequency slots that are not over-crowded.

\item \textit{Geographic routing}: Position information is the main input for geographic routing, in which a wireless node attempts to send the data packet to the destination node in a multi-hop fashion. This can be achieved, if each node, that holds the packet, forwards the packet to the closest node (e.g., measured by Euclidean distance) to the destination. Consequently, the latency of this routing method depends on the position accuracy.

\item \textit{Autonomous driving}: Position awareness for autonomous vehicles is crucial since most of the vehicle's actions depend on its current location (e.g., computing minimum distance paths, or avoiding collisions). Since this problem requires extremely high accuracy and robustness, the available measurements from DM-MIMO should be fused with all other available information (especially, from the GPS and the road maps).
\ei

However, since current positioning systems are not ideal, and require a trade-off between different metrics, there are still many issues to be addressed. We discuss here the three most important issues, with a focus on fingerprinting methods.
\bi
\item \textit{Performance in dynamic environments}: It is well known that fingerprinting methods may fail in dynamic environments, i.e., if the positions of reflecting/diffracting surfaces frequently change. In principle, this problem can be solved with online training, assuming that enough access points (``sniffing devices'') with known positions are deployed. Alternatively, the online training can be perfomed using a mobile vehicle with GPS capability.

\item \textit{Communication overhead}: The positioning system is expected to use the RSS fingerprints obtained at predefined locations, which would affect the communication overhead. That means that the frequency of the training (especially, if it is online), and the number of training points should be carefully chosen, to avoid an excess in communication cost. If this is not possible, some efficient data compression technique should be developed.

\item \textit{Latency}: One of the main problems of fingerprinting methods is the relatively high computational complexity. With nowadays microprocessors, this would not create a latency issue in most situations. However, in certain scenarios, in which many training points are required, or if the position estimates are required very frequently, the latency becomes an important problem. Therefore, the development of faster fingerprinting algorithms is an important challenge.
\ei

\section{Conclusions}\label{sec:conc}
In this paper, we discussed possible positioning techniques for DM-MIMO. We argued that FP-based positioning is an appropriate solution compared with standard (triangulation and trilateration) methods. We also provided a solution for FP-based positioning based on GPR. This method is able to model arbitrary nonlinear relationships and provide probabilistic outputs, in contrast to other FP methods.
However, there remain many challenges that should be addressed. The most critical problem is to ensure good performance in dynamic environments, while keeping the latency and the communication overhead acceptable. In summary, positioning with massive MIMO is an unexplored problem, with a lot of possibilities for further research.

\section*{Acknowledgment}
This work was supported by the project Cooperative Localization (CoopLoc) funded by the Swedish Foundation for Strategic Research (SSF), and Security Link.

\balance
\footnotesize %
\bibliographystyle{IEEEbib}
\bibliography{refsPaper}

\end{document}